\documentclass[conference]{IEEEtran}
\IEEEoverridecommandlockouts
\usepackage{cite}
\usepackage{amsmath,amssymb,amsfonts}
\usepackage{algorithm}
\usepackage{algorithmic}
\usepackage{array}
\usepackage{booktabs}
\usepackage{graphicx}
\usepackage{subfigure}
\usepackage{textcomp}
\usepackage{xcolor}
\usepackage{graphicx}
\usepackage{tablefootnote}
\usepackage{multirow}
\usepackage{marvosym}
\usepackage[T1]{fontenc}
\usepackage{amsmath,graphicx,amsfonts,svg,float}
\usepackage{booktabs,multirow,stfloats}
\usepackage[colorlinks,
            linkcolor=blue,
            anchorcolor=blue,
            citecolor=blue]{hyperref}

\def\BibTeX{{\rm B\kern-.05em{\sc i\kern-.025em b}\kern-.08em
    T\kern-.1667em\lower.7ex\hbox{E}\kern-.125emX}}
\begin{document}

\title{\textsc{MAIN-VC}: Lightweight Speech Representation Disentanglement for One-shot Voice Conversion
}


\author{\IEEEauthorblockN{Pengcheng Li$^{1,2\ddag}$, Jianzong Wang$^{1\ddag}$ \thanks{$\ddag$ \  Equal contribution. }, Xulong Zhang$^{1\textsuperscript{\Letter}}$ \thanks{$\textsuperscript{\Letter}$ Corresponding author.}, Yong Zhang$^{1}$, Jing Xiao$^{1}$, Ning Cheng$^{1}$}
\IEEEauthorblockA{\textit{$^{1}$Ping An Technology (Shenzhen) Co., Ltd.}\\\textit{$^{2}$University of Science and Technology of China}}
\IEEEauthorblockA{
\texttt{lipengcheng@ustc.edu, jzwang@188.com, zhangxulong@ieee.org, }
\\ \texttt{yzhang3272-c@my.cityu.edu.hk, \{xiaojing661, chengning211\}@pingan.com.cn}
}}

\maketitle

\begin{abstract}
One-shot voice conversion aims to change the timbre of any source speech to match that of the unseen target speaker with only one speech sample. Existing methods face difficulties in satisfactory speech representation disentanglement and suffer from sizable networks as some of them leverage numerous complex modules for disentanglement. In this paper, we propose a model named \textsc{MAIN-VC} to effectively disentangle via a concise neural network. The proposed model utilizes Siamese encoders to learn clean representations, further enhanced by the designed mutual information estimator. The Siamese structure and the newly designed convolution module contribute to the lightweight of our model while ensuring performance in diverse voice conversion tasks. The experimental results show that the proposed model achieves comparable subjective scores and exhibits improvements in objective metrics compared to existing methods in a one-shot voice conversion scenario. The code is available at \url{https://github.com/PecholaL/MAIN-VC}.
\end{abstract}


\section{Introduction}
Voice conversion (VC) is the style transfer task in the field of speech. It modifies the source speech's style to the target speaker's speaking style. Voice conversion can be applied in smart devices, entertainment industry, and even privacy security area \cite{Yuan/2022/Deid,Lal2020evaluate}.

Early voice conversion relies heavily on parallel training data \cite{Sisman/2021/An}, while parallel data collecting and frame-level alignment procedures consume much time. So researchers have turned attention to non-parallel voice conversion, enabling the possibility of many-to-many and any-to-any conversion tasks. Inspired by image style transfer in computer vision, a series of methods based on generative adversarial network (GAN) have been applied to voice conversion \cite{Kaneko/2018/CycleGAN-VC,Kaneko/2021/StarGAN-VC2,Li/2021/star,Chen/2022/effi}. These GAN-based VC methods treat each speaker as a different domain and consider voice conversion as migration across speaker domains. But GAN training is unstable and poses difficulties in convergence. 

Recently, disentanglement-based VC \cite{luong2021many,xiao2022dgc,liu2023automatic} has been an area of active research for its flexibility in any-to-any voice conversion. The disentanglement-based VC method is built on the assumption that speech composition contains both speaker-dependent information (\textit{e.g.} timbre and speaking style) and speaker-independent information (\textit{e.g.} content). Generally, speaker-dependent information is time-invariant, remaining constant in one utterance, whereas speaker-independent information is time-variant. This allows neural networks to separately learn representations of content and speaker information from the speech. The disentanglement-based VC models typically adopt an autoencoder structure, integrating approaches like bottleneck features \cite{Qian/2019/AutoVC}, instance normalization \cite{Chou/2019/One-Shot}, vector quantization \cite{Wu/2020/One-shot}, \textit{etc.} into encoders to extract desired speech representations. The training usually involves reconstruction of Mel-spectrogram. During training, the same Mel-spectrogram segments are fed into the decoupling encoders to obtain various speech representations, then the decoder integrates the extracted speech representations and reconstruct the Mel-spectrogram. After the training is completed, the content information from the source speech and the speaker information from the target speech is decoupled respectively and then fused for synthesizing the converted voice, achieving the transferring of speaker style. 

However, the disentanglement of speech representation is complicated \cite{yuan2021improving}, how to effectively eliminate unnecessary information while preventing damage to the wanted information that needs to be disentangled, and how to extract clean representations without including other information, are challenging aspects to master. Most of the previous voice conversion models suffer unsatisfactory disentangling performance. Besides the dependence of some models on the fine-tuned bottleneck \cite{Qian/2019/AutoVC,Qian/2020/Unsupervised}, the reasons for deficient disentanglement include that different encoders disentangle independently during the representation learning process, lacking communication with the others, resulting in the existence of redundant information in disentangled representations. Moreover, most of these VC models are sizable, as some of them improve disentangling capability through stacking various network modules, making it difficult to deploy the models on low-computing mobile devices. 

In this paper, we propose a model based on speech representation disentanglement, named \textbf{M}utual information enhanced \textbf{A}daptive \textbf{I}nstance \textbf{N}ormalization \textbf{VC} (\textbf{\textsc{MAIN-VC}}). To enhance the disentangling ability of the encoders, we integrate the Siamese structure as well as data augmentation into the speaker representation learning module. An easy-to-train constrained mutual information estimator is introduced to build the communication between the learning processes of content and speaker information, which reduces the overlap between unrelated representations. With the assistance of these modules, the disentangling ability of the encoders no longer rely on complicated networks. This makes it possible for model lightweight and deployment on low-computing devices. The contributions of this paper can be summarized as follows:
\begin{itemize}
    \item We introduce the \textit{speaker information learning module} which reduces the interference of time-varying information to extract clean speaker embeddings.
    \item We propose the \textit{constrained mutual information} estimator restricted by upper and lower bounds and leverage it to enhance disentangling capability, contributing to improve the quality of one-shot VC.
    \item We design the proposed model to be lightweight while ensuring the quality of voice conversion. A convolution block called APC with a low parameter count and large receptive field is also introduced to commit to lightweight.
\end{itemize}

\section{Related Works}
\subsection{Speech Representation Disentanglement}
Speech representation disentanglement (SRD) decomposes speech components into diverse representations. It has been used in speaker recognition \cite{liu2023disentangling}, speaker diarization \cite{mun2023eend}, speech emotion recognition \cite{latif2021survey}, speech enhancement \cite{hou2022learning}, automatic speech synthesis \cite{wang2023generalizable}, \textit{etc}. 

SRD also conveniently provides an opportunity for the purpose of voice conversion, altering the speaker identity information in speech to achieve conversion. The voice conversion methods based on SRD have gained widespread attention in recent years. \textsc{AutoVC} \cite{Qian/2019/AutoVC} leverages an autoencoder with a carefully tuned bottleneck to remove speaker information. A pre-trained speaker encoder is utilized to provide speaker identity information. \textsc{SpeechSplit} \cite{Qian/2020/Unsupervised} and \textsc{SpeechSplit2.0} \cite{Chan/2022/SpeechSplit2} design multiple encoders with signal processing to decouple finer-grained speech components such as rhythm, content and pitch. Another series of works \cite{Chou/2019/One-Shot,Yen/2021/Again} removes time-invariant speaker information through instance normalization to get content representation, then extracted speaker representation is fused into the content representation in an adaptive manner. Vector quantization (VQ) is also leveraged to complete speech representation disentanglement. In vector quantization-based methods \cite{Wu/2020/VQVCPlus,Wang/2021/Vqmivc,Yang/2022/Streamable,Tang/2022/avqvc}, latent representation is quantized into discrete codes, which are highly related to phonemes, and the differences between the discrete codes and the vectors before quantization are considered as the speaker information.

SRD-base VC method shows its flexibility in a one-shot voice conversion scenario but places high demands on the model's disentangling ability.

\subsection{Neural Mutual Information Estimation}
Mutual information (MI) measures the degree of correlation or inclusion between two variables. The calculation of MI generally requires calculating the joint distribution and marginal distribution of the variables. However, high-dimensional random variables from neural networks often own high complexity and complicated distribution. Only samples can be obtained while distribution is difficult to derive. It is difficult to accurately obtain the mutual information between them. Therefore, neural MI estimation is proposed to address this issue. 

MINE \cite{ishmael/2018/mine} treats MI as the Kullback-Leibler divergence (KL divergence) between the joint and marginal distributions, and sets a function to fit this KL divergence. The fitting process is completed via a trainable neural network. InfoNCE \cite{oord2018representation} maximizes the MI between a learned representation and the data and estimates the lower bound of MI based on noise contrastive estimation \cite{gutmann2010noise}. 
Representative works on mutual information upper bound estimation include L1Out \cite{ben2019on} and CLUB \cite{Cheng/2020/club}. The former leverages the Monte Carlo approximation to approximate the marginal distribution and then derives the upper bound of MI. CLUB \cite{Cheng/2020/club} fits the conditional probability between two variables. It designs positive sample pairs and negative sample pairs to construct a contrastive probability log ratio between two conditional distributions. When the approximate condition probability is close enough to the truth (\textit{i.e.} the KL divergence is small enough), the fitted mutual information gradually approaches the true MI from above, thus obtaining the upper bound of MI. 

In practical applications, the estimation of the lower bound of the MI is maximized when it is necessary to strengthen the correlation between two types of variables, while the upper bound of the MI is minimized when reducing the correlation between variables.

\begin{figure*}[t]
\centering
\centerline{\includegraphics[width=0.85\textwidth]{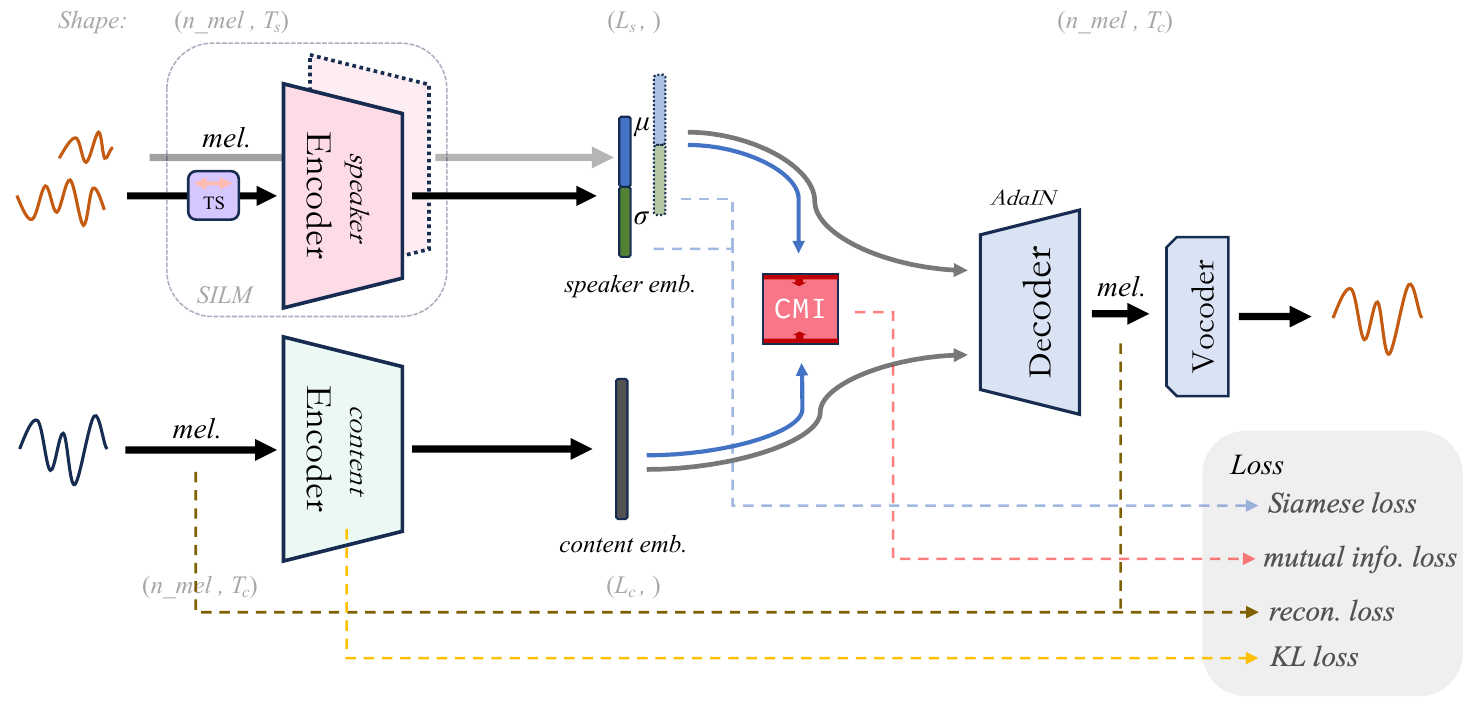}}
\caption{Architecture of \textsc{MAIN-VC} and the training objectives.}
\label{fig:frame}
\end{figure*}

\section{Methodology}
\subsection{Overall Architecure of \textsc{MAIN-VC}}
The architecture of \textsc{MAIN-VC} is depicted in Fig. \ref{fig:frame}. \textsc{MAIN-VC} adopts an instance normalization- (IN-)based method for speech style transferring. A speaker information learning module consists of Siamese encoders with a data augmentation module built to obtain clean speaker representation, while the constrained mutual information estimator between speaker representation and content representation is leveraged to enhance the disentanglement. The pre-trained vocoder only appears during the inference stage. The constrained mutual information estimator and the Siamese encoder both take part in the training stage to assist in the training of the content encoder and the speaker information learning module. They are not involved in the conversion process during the inference stage after the training is completed.

\subsection{IN-based Speech Style Transfer}
When given frame-level linguistic feature $Z$, the content encoder and speaker representation learning module extract content representation $z_C$ and speaker representation $z_S$ respectively. To mitigate the impact of time-invariant information like timbre on the extraction of content representation, the content encoder removes speaker information from the mid feature map $M$ (processed from linguistic feature $Z$) through instance normalization without affine transformation (denoted as $\textrm{IN}_{woAT}$) as Eq. \ref{eq:in} shows,
\begin{equation}
\begin{split}
\label{eq:in}
& \mu(M_{[c]})=\frac{1}{W}\sum_{w=1}^{W}M_{[c]}^{[w]} \\
& \sigma(M_{[c]})=\sqrt{\frac{1}{W}\sum_{w=1}^{W}(M_{[c]}^{[w]}-\mu(M_{[c]}))^2} \\
& \textrm{IN}_{woAT}(M_{[c]}^{[w]})=\frac{M_{[c]}^{[w]}-\mu(M_{[c]})}{\sigma(M_{[c]})}
\end{split}
\end{equation}
where $M_{[c]}^{[w]}$ represents the $w$-th element of the $c$-th channel of feature map $M$. Content representation $z_C$ can be extracted from $IN_{woAT}(M_{src})$ where $M_{src}$ is the feature map obtained from the linguistic feature $Z_{src}$ of the source speech, while the global speaker representation $z_S$ is the tuple $(\alpha,\beta)$ obtained from the linguistic feature of the target speech $Z_{tgt}$ via speaker encoder. Once we collect the content representation $z_C$ and speaker representation $z_S$, the decoder integrates speaker information via its internal adaptive instance normalization (AdaIN) \cite{Huang/2017/Arbitrary} layers:
\begin{equation}
\label{eq:adain}
\textrm{AdaIN}({M_C}_{[c]}^{[w]}, {z_S}_{[c]})= \beta_{[c]}\cdot\frac{{M_C}_{[c]}^{[w]}-\mu({M_C}_{[c]})}{\sigma({M_C}_{[c]})}+\alpha_{[c]}
\end{equation}
where $M_C$ is the feature map obtained from content representation $z_C$ via the internal blocks of the decoder, and ${z_S}_{[c]}$ (composed of $\alpha_{[c]}$ and $\beta_{[c]}$) denotes the $c$-th channel of the speaker representation. Essentially, AdaIN layers adaptively adjust the affine parameters of IN according to speaker information to align the channel-wise statistics of content with those of speaker style to achieve style transferring. Extracting continuous speaker representations from arbitrary target utterances enables the model applicability in any-to-any VC scenarios.

\subsection{Speaker Information Learning Module}
\label{ssec:siamese}
Although SRD-based VC models do not require parallel corpus for training, typically using multi-speaker datasets where the same speaker has multiple utterances, to fully utilize this characteristic of multi-speaker datasets, we generalize the time-invariance within utterance to the \textit{utterance-invariance} across utterance from the same speaker. 

In our proposed method, the speaker information learning module (SILM) based on utterance-invariance consists of Siamese encoders (denoted as $Enc_S$ and $Enc_S'$) and a time shuffle unit (TS). TS processes the input data (\textit{i.e.} target speech) before it enters $Enc_S$, it rearranges the order of frame-level segments along time dimension to disrupt time-variant information (\textit{i.e.} content information) while preserving time-invariant information (\textit{i.e.} speaker information). During the training stage, the same linguistic feature $Z$ is fed into both $Enc_S$ of SILM and content encoder $Enc_C$, then the decoder $Dec$ performs reconstruction with the outputs of the encoders as shown in Eq. \ref{eq:recon}. Different from the previous works, the parallel time-variant content information is completely disrupted in the proposed method.
\begin{equation}
\label{eq:recon}
    \hat{Z}=\mathbf{Dec}(\mathbf{Enc_C}(Z), \mathbf{Enc_S}(\mathbf{TS}(Z)))
\end{equation}

Linguistic feature $Z'$ of another utterance from the same speaker is fed into $Enc_S'$ of SILM after reconstruction. Based on the utterance-invariance of speaker information, we use different utterances from the same speaker for the consideration that speech inflection and recording conditions affect the extraction of speaker representation. The Siamese loss shown in Eq. \ref{eq:sialoss} helps to improve the cosine similarity between these two speaker representations: 
\begin{equation}
\label{eq:sialoss}
    \mathcal{L}_{Siamese}=1-\mathit{cos}(\mathbf{Enc_S}(\mathbf{TS}(Z)), \mathbf{Enc_S'}(Z'))
\end{equation}

In the inference stage, the Siamese structure is disregarded, since only one utterance from the target speaker is required.

\subsection{Mutual Information Enhanced Disentanglement}
\label{ssec:mi}

We design the constrained mutual information (CMI) estimator with upper and lower bounds to estimate mutual information. Estimating MI between complex random variables is challenging and always unstable to train in practice, simultaneously estimating both upper and lower bound of it and constraining them mutually help to enhance the MI estimator's robustness and prediction accuracy. The estimation of the lower bound can serve as guidance and a baseline for the upper bound estimation, resulting in a more trainable mutual information estimator, CMI. In \textsc{MAIN-VC}, the upper bound of MI between content representation $z_C$ and speaker representation $z_S$ is derived from the variational contrastive log-ratio upper bound (vCLUB) \cite{Cheng/2020/club} as:

\begin{equation}
\label{eq:vclub}
\begin{split}
\textrm{I}_{\mathit{upper}}(z_C; z_S) = & \mathbb{E}_{\mathrm{\mathit{P}(z_C, z_S)}}[\log(\mathit{Q_\theta}(z_C|z_S))] \\
& - \mathbb{E}_{\mathit{P}(z_C)}\mathbb{E}_{\mathit{P}(z_S)}[\log(\mathit{Q_\theta}(z_C|z_S))]
\end{split}
\end{equation}
where $\mathit{P}(z_C|z_S)$ is the actual conditional distribution, while $\mathit{Q_\theta}(z_C|z_S)$ represents its variational estimation obtained via a trainable network. The lower bound of MI can be estimated through MINE \cite{ishmael/2018/mine}, which employs a function $\textrm{T}_\theta$ with trainable parameters $\theta$ to approximate the KL divergence between two distributions, \textit{i.e.} the joint distribution $\mathit{P}(z_C, z_S)$ and the product of the two marginal distributions $\mathit{P}(z_C)\otimes \mathit{P}(z_S)$, in order to predict the lower bound of $I(z_C;z_S)$, as:

\begin{equation}
\label{eq:mine}
\begin{split}
\textrm{I}_{\mathit{lower}}(z_C; z_S) = & \mathbb{E}_{\mathrm{\mathit{P}(z_C, z_S)}}[\textrm{T}_\theta(z_C, z_S)] \\
& - \log(\mathbb{E}_{\mathit{P}(z_C)\otimes \mathit{P}(z_S)}[\mathit{e}^{\textrm{T}_\theta(z_C, z_S)}])
\end{split}
\end{equation}

CMI is trained using the learning losses of the upper and lower bound estimation networks, as well as the gap between their estimated MI values. We minimize the estimated upper bound for further disentanglement of the speech representations. When $N$ utterances are given, and the length of content representation extracted from each cropped utterance is $L$, then CMI estimates MI loss as:

\begin{equation}
\label{eq:miloss}
    \mathcal{L}_{MI}=\frac{1}{N^{2}L}{\sum\limits_{m=1}^{N}{\sum\limits_{n=1}^{N}{\sum\limits_{l=1}^{L}{\log \frac{{Q_\theta }\left( {{{\mathit{z_C}}}_{m,l}}|{{\mathit{z_S}}_{n}} \right)}{ {Q_\theta}\left( {{{\mathit{z_C}}}_{n,l}}|{{\mathit{z_S}}_{n}} \right)}}}}} 
\end{equation}

It's worth noting that SILM's enhancement of speaker representation disentangling can also facilitate CMI's effect on content disentangling implicitly, as the cleaner speaker representation obtained from SILM, the more accurate content-irrelevant information that CMI needs to eliminate during content disentangling.

\subsection{Model Lightweighting}
\label{ssec:lightw}

\begin{figure}[t]
\centering
\setlength{\abovecaptionskip}{-0.22cm}
\includegraphics[width=.49\textwidth]{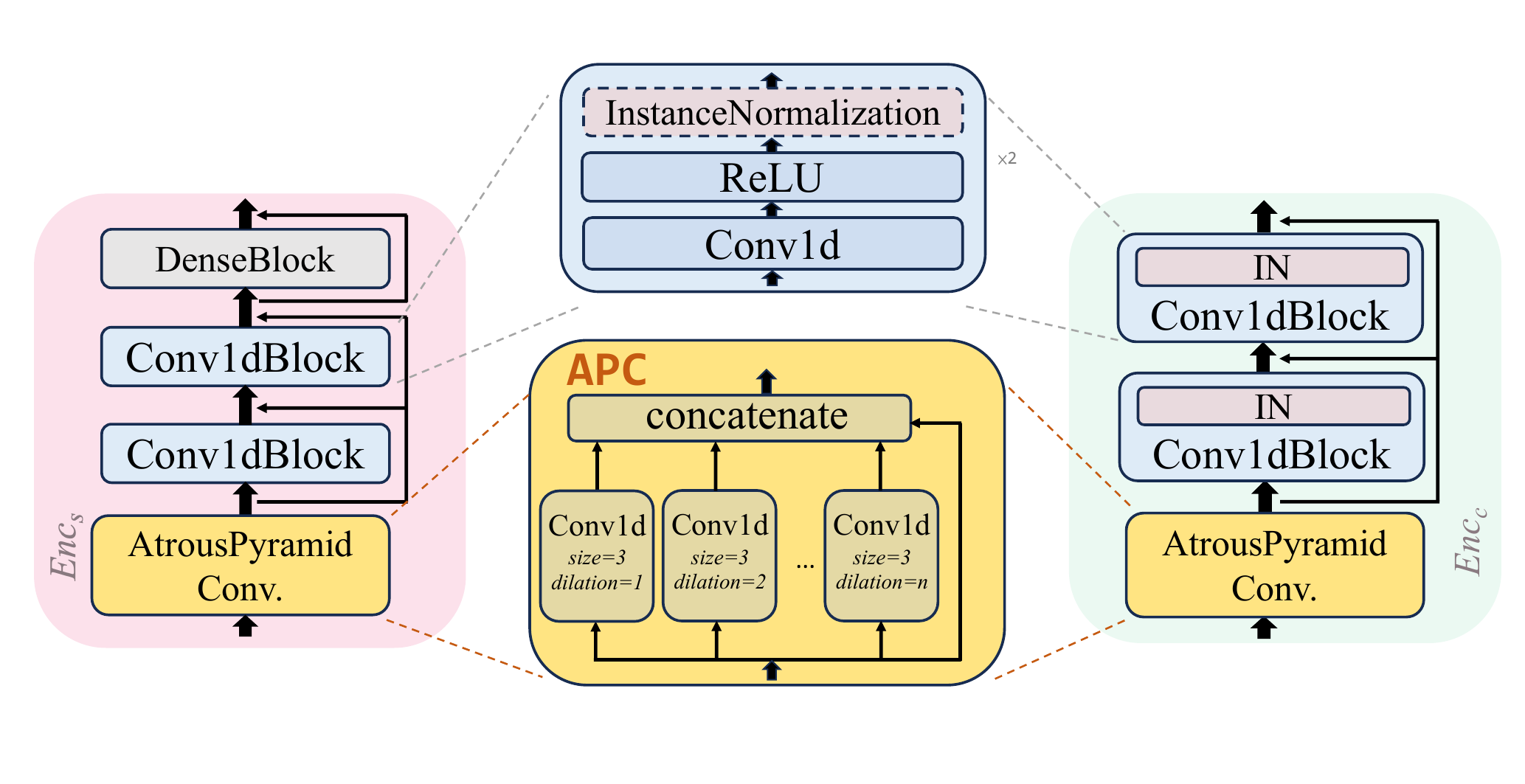}
\caption{The structure of the encoders and the details of APC.}
\label{fig:arch}
\end{figure}

Inspired by atrous spatial pyramid pooling (ASPP) \cite{Liang/2018/Deeplab}, we design a convolution block named atrous pyramid convolution (APC) as shown in Fig. \ref{fig:arch}. APC contains a series of small kernels ($kernel\_size=3$ in \textsc{MAIN-VC}) with different dilation factors \cite{fisher2017dilated} and a residual-connect concatenating the convolution results, replacing the sizable convolution block in previous works. The previous convolution block consists of a set of kernels of different sizes (size from $1$ to $8$ in the baseline method \cite{Chou/2019/One-Shot}) which contain large kernels. APC not only reduces the parameter count significantly but also expands the receptive field. 

What's more, parameter sharing in the Siamese structure of SILM also prevents the network from growing. CMI only appears during training and makes no time consumption during inference. We adjust the layers and channels of the encoders to strike a balance between performance and model lightweight.

\subsection{Loss Function}
\label{ssec:lossfunc}
The loss function of MAIN-VC contains reconstruction loss, KL divergence loss, MI loss and Siamese loss. Reconstruction loss is the $\textrm{L1}$ loss between the source log-Mel-spectrogram $Z \in \mathbb{R}^{M \times T}$ and the output of decoder $\hat{Z} \in \mathbb{R}^{M \times T}$(as shown in Eq. \ref{eq:recon} and Eq. \ref{eq:reconloss}). The KL divergence between the posterior distribution $\mathit{P}(z_C|Z)$ and the predefined prior distribution $\mathcal{N}(0,I)$ serves as the KL loss, shown in Eq. \ref{eq:losskl}. 

\begin{equation}
\label{eq:reconloss}
\mathcal{L}_\textit{recon}=\Vert Z - \hat{Z} \Vert^1_1
\end{equation}

\begin{equation}
\setlength\abovedisplayskip{-0.2cm}
\setlength\belowdisplayskip{0.4cm}
\mathcal{L}_{KL} = \mathbb{E}_{P(Z)} [\Vert {\mathbf{Enc_C}(Z)}^2 \Vert^2_2]
\label{eq:losskl}
\end{equation}

Siamese loss and MI loss are already defined in Eq. \ref{eq:sialoss} and Eq. \ref{eq:miloss} respectively. Finally, the total loss combines the aforementioned losses with weights $\lambda_1$, $\lambda_2$ and $\lambda_3$, as:
\begin{equation}
\label{eq:loss}
\mathcal{L}=\mathcal{L}_\textit{recon}+\lambda_1\mathcal{L}_{KL}+\lambda_2\mathcal{L}_\textit{Siamese}+\lambda_3\mathcal{L}_{MI}
\end{equation}

\section{Experiments and Results}
\label{sec:experiments}

\subsection{Experimental Setup}
\label{ssec:setup}

To evaluate \textsc{MAIN-VC}, a comparative experiment and an ablation study are conducted on VCTK \cite{vctk2016} corpus. 19 speakers are randomly selected as unseen speakers for one-shot conversion tasks, and their speech is excluded during training. All the utterances are resampled to 16,000 Hz during the experiment, and the Mel-spectrograms are computed through short-time Fourier transform for training. During the training set sampling, each utterance is randomly cropped into several 128-frame segments. Two segments from different utterances spoken by the same speaker are then paired to form a training data sample, where one is used for reconstruction, and the other is fed into the Siamese encoder of SILM.

It's worth noting that \textsc{MAIN-VC} requires at least two forward and backward passes in each training iteration, as the CMI module needs extra training. In other words, the training process involves the following two steps in each iteration: 
\begin{itemize}
  \item [1)] 
  Training the integrated CMI network, the gap between upper and down bounds is combined with the learning losses. This step can be repeated several times in one iteration (five times in our experiment).
  \item [2)]
  Computing the four loss components to obtain the total for the training of the entire model (one forward and backward propagation, excluding the pre-trained vocoder). 
\end{itemize}
\textsc{MAIN-VC} is trained with the Adam optimizer \cite{kingma2015adam} with $\beta_1=0.9$, $\beta_2=0.99$, $\epsilon=10^{-6}$, $\textit{lr}=1\times10^{-4}$. Another Adam optimizer with $\textit{lr}=2\times10^{-4}$ is built to train the CMI network. During the inference stage, a pre-trained WaveRNN \cite{nal2018efficient} is used as the vocoder to generate waveform from Mel-spectrogram. We design the weights in the total loss function Eq. \ref{eq:loss} based on the numerical characteristics of each component, $\lambda_1$ and $\lambda_3$ gradually increase to 1 as the number of iterations progresses, while $\lambda_2$ is set to 1. We compare our proposed models with AdaIN-VC \cite{Chou/2019/One-Shot}, \textsc{AutoVC} \cite{Qian/2019/AutoVC}, VQMIVC \cite{Wang/2021/Vqmivc} and LIMI-VC \cite{Huang/2023/Limivc}.

\begin{table*}[t]
    \vspace{-1em}
  \centering
  \setlength{\abovecaptionskip}{0pt}
  \setlength{\belowcaptionskip}{10pt}
  \caption{Comparison of different methods for many-to-many and one-shot VC (with 95\% confidence interval).}
  \label{tab:res}
    \begin{tabular}{p{2.8cm}<{\centering}cccccc}
    \toprule
    \multirow{2}{*}{\textbf{VC Method}}&
    \multicolumn{3}{c}{\textbf{Many-to-Many}}&\multicolumn{3}{c}{\textbf{One-Shot}}\cr
    \cmidrule(lr){2-4} \cmidrule(lr){5-7}
    & MCD$\downarrow$ & MOS$\uparrow$ & VSS$\uparrow$ 
    & MCD$\downarrow$ & MOS$\uparrow$ & VSS$\uparrow$\cr
    \midrule
    AdaIN-VC \cite{Chou/2019/One-Shot}
             & 7.64 $\pm$ 0.24 & 3.11 $\pm$ 0.13 & 2.82 $\pm$ 0.19
             & 7.38 $\pm$ 0.14 & 3.04 $\pm$ 0.21 & 2.45 $\pm$ 0.16
    \cr
    \textsc{AutoVC} \cite{Qian/2019/AutoVC}
           & 6.68 $\pm$ 0.21 & 2.76 $\pm$ 0.16 & 2.45 $\pm$ 0.23
           & 8.08 $\pm$ 0.15 & 2.68 $\pm$ 0.17 & 2.39 $\pm$ 0.14
    \cr
    VQMIVC \cite{Wang/2021/Vqmivc}
           & 6.24 $\pm$ 0.13 & 3.20 $\pm$ 0.14 & 3.32 $\pm$ 0.12
           & 5.59 $\pm$ 0.10 & 3.16 $\pm$ 0.18 & 3.02 $\pm$ 0.18
    \cr
    LIMI-VC \cite{Huang/2023/Limivc}
           & 7.27 $\pm$ 0.17 & 3.12 $\pm$ 0.13 & 3.01 $\pm$ 0.26
           & 7.88 $\pm$ 0.18 & 2.97 $\pm$ 0.22 & 2.74 $\pm$ 0.24
    \cr
    \midrule
    \textbf{\textsc{MAIN-VC}$_{large}$} & 5.17 $\pm$ 0.14 & 3.41 $\pm$ 0.21 & 3.35 $\pm$ 0.12
           & 5.31 $\pm$ 0.09 & 3.33 $\pm$ 0.17 & 3.23 $\pm$ 0.17
    \cr
    \textbf{\textsc{MAIN-VC}} & 5.28 $\pm$ 0.11 & 3.44 $\pm$ 0.12 & 3.25 $\pm$ 0.16
                     & 5.42 $\pm$ 0.13 & 3.24 $\pm$ 0.18 & 3.29 $\pm$ 0.11
    \cr
    \bottomrule
    \end{tabular}
    \vspace{-1.2em}
\end{table*}

\subsection{Metrics and Measurement}
The following subjective and objective metrics are used to evaluate the performance of different voice conversion models. \\
\textbf{Mean opinion score (MOS)} is a common subjective metric which assesses the naturalness of speech, with scores ranging from 1 to 5, where higher scores indicate higher speech quality. \\
\textbf{Voice similarity score (VSS)}, a subjective metric, evaluates the similarity between the converted speech and the ground-truth speech from the target speaker. Higher scores indicate higher similarity, representing better VC performance. \\
\textbf{Mel-cepstral distortion (MCD)} is an objective metric to measure the difference between the converted speech and the target ground-truth speech in the aspect of Mel-cepstral.

Both many-to-many and one-shot (any-to-any) VC are conducted while refraining from using parallel data during the inference stage. 4 pairs of source/target speech are input into each VC model for evaluation in two scenarios respectively. Then 8 participants are invited to score the speech samples. 

For the measurement of lightweight, we choose \textbf{parameter count} and \textbf{inference time cost} as the metrics. The measurements for both do not include the vocoders, as the vocoder in each model can be freely replaced with an appropriate one. The inference time is measured and averaged using the same utterance pairs with each model on the same device ($1 \times$ Tesla V100).

\begin{figure*}[htb]
    \centering

    \setlength{\abovecaptionskip}{-0.02cm}
    \includegraphics[width=0.97\textwidth]{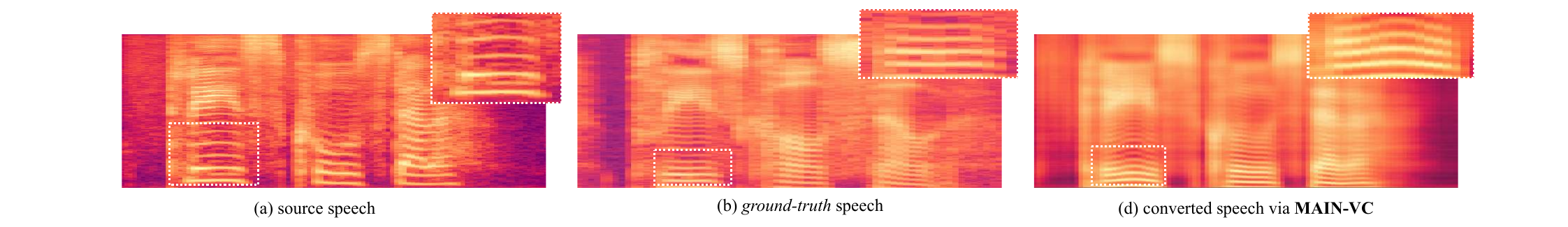}
    \caption{The Mel-spectrogram samples of the one-shot VC task "\textit{Please call Stella}".}
    \label{fig:mel}
\end{figure*}

\subsection{Experimental Results}
Table \ref{tab:res} showcases the comparison of \textsc{MAIN-VC} and baseline methods in different voice conversion scenarios. The proposed non-lightweight model (denoted as \textsc{MAIN-VC}$_{large}$) is also taken into consideration. For many-to-many VC, \textsc{MAIN-VC} achieves comparable performance with baseline methods. \textsc{MAIN-VC} outperforms other methods in one-shot VC, especially on the objective metric. As for the lightweight of each model, the comparison of the parameter count and inference time cost (exclude vocoder) is presented in Table \ref{tab:light}. \textsc{MAIN-VC} reaches $59\%$ of reduction in the parameter count and $33\%$ of reduction in the inference cost, compared to the lightest baseline methods respectively. The proposed model has an obvious advantage in parameter count thanks to the proposed APC module and the parameter-sharing strategy, also inferences less costly due to its lightweight network. Overall, \textsc{MAIN-VC} strikes a balance between lightweight design and performance, obtaining a light network structure while ensuring the quality of voice conversion.

Fig. \ref{fig:mel} showcases Mel-spectrograms of utterances in a one-shot VC task, including the source speech, ground-truth speech (\textit{i.e.} the same content to source speech spoken by the target speaker, \textbf{not} the target speech in the inference), converted speeches via \textsc{MAIN-VC}. The details of the spectrograms show that the source speaker and target speaker own different acoustic characteristics (as in Fig. \ref{fig:mel}(a) and Fig. \ref{fig:mel}(b), the distributions and shapes of the harmonic from different speakers are quite different). \textsc{MAIN-VC} can extract the feature of the target speaker and construct the converted Mel-spectrogram that is close to the ground-truth. \textsc{MAIN-VC} is also capable of performing cross-lingual voice conversion (XVC), another one-shot VC scenario, where the source and target utterances are in different languages. We conduct XVC on AISHELL \cite{aishell2018} (Mandarin speech corpus) and VCTK \cite{vctk2016} (English speech corpus). More audio samples are available at \url{https://largeaudiomodel.com/main-vc/}.



\begin{table}[htbp]
\setlength{\abovecaptionskip}{-0.30cm}
\small
    \vspace{-1em}
   \centering
   \caption{Comparison of parameter count and inference time cost \\(\textit{w/o} vocoder).} 
    \label{tab:light}
    \begin{tabular}{ccc} \\
    \toprule
    \textbf{VC Method} & \textbf{Param. Count}$\downarrow$ & \textbf{Infer. Cost}$\downarrow$
    \cr
    \midrule
    AdaIN-VC \cite{Chou/2019/One-Shot} & 9.04M & 100\% 
    \cr
    \textsc{AutoVC} \cite{Qian/2019/AutoVC} & 28.27M & 50.94\% 
    \cr
    VQMIVC \cite{Wang/2021/Vqmivc} & 29.20M & 145.28\% 
    \cr
    LIMI-VC \cite{Huang/2023/Limivc} & 3.19M & 49.06\% 
    \cr
    \midrule
    $\textbf{\textsc{MAIN-VC}}$ & 1.31M & 22.64\% 
    \cr
    \bottomrule
    \end{tabular}
    \vspace{-0.5em}
\end{table}

\subsection{Ablation Study}
To validate the effects of SILM and CMI on disentanglement, we conduct ablation experiments. We set up the following models:
\begin{itemize}
  \item [a)] \textsc{MAIN-VC}: the proposed model.
  \item [b)] M1 (\textit{w/o} CMI): remove the constrained mutual information estimator from the proposed method.
  \item [c)] M2 (\textit{w/o} CMI$_{lower}$): only use the estimator provided by CLUB \cite{Cheng/2020/club} to assess the upper of mutual information.
  \item [d)] M3 (\textit{w/o} Siamese encoder): remove the Siamese structure and time shuffle unit in SILM. 
  
\end{itemize}

Resemblyzer is employed for fake speech detection to assess the conversion quality. Resemblyzer simultaneously scores the fake (\textit{i.e.} the outputs of VC model in the experiment) and real utterances of the target speaker after learning the characteristics of the target speaker from another 6 bonafide utterances. A higher score indicates more similar timbre and better speech quality. The results are presented in Table \ref{tab:ablation}. CMI and SILM contribute to \textsc{MAIN-VC}'s synthesis of converted speech that is more confusing for fake detection, and the lower bound of CMI also shows its beneficial effect on the converted speech.

\begin{table}[htbp]
\setlength{\abovecaptionskip}{-0.0cm}
\small
    \vspace{-1em}
   \centering
   \caption{Comparison for ablation study.} 
    \label{tab:ablation}
    \begin{tabular}{cc} \\
    \toprule
    \textbf{Method} & \textbf{Detection Score}$\uparrow$
    \cr
    \midrule
    \textbf{\textsc{MAIN-VC}} & 0.74 $\pm$ 0.01
    \cr
    M1 (\textit{w/o CMI}) & 0.62 $\pm$ 0.04  
    \cr
    M2 (\textit{w/o CMI$_{lower}$}) & 0.69 $\pm$ 0.03
    \cr
    M3 (\textit{w/o Siamese Enc.}) & 0.70 $\pm$ 0.03  
    \cr
    \bottomrule
    \end{tabular}
    \vspace{-0.1em}
\end{table}

\begin{figure}[t]
\centering
\includegraphics[width=.48\textwidth]{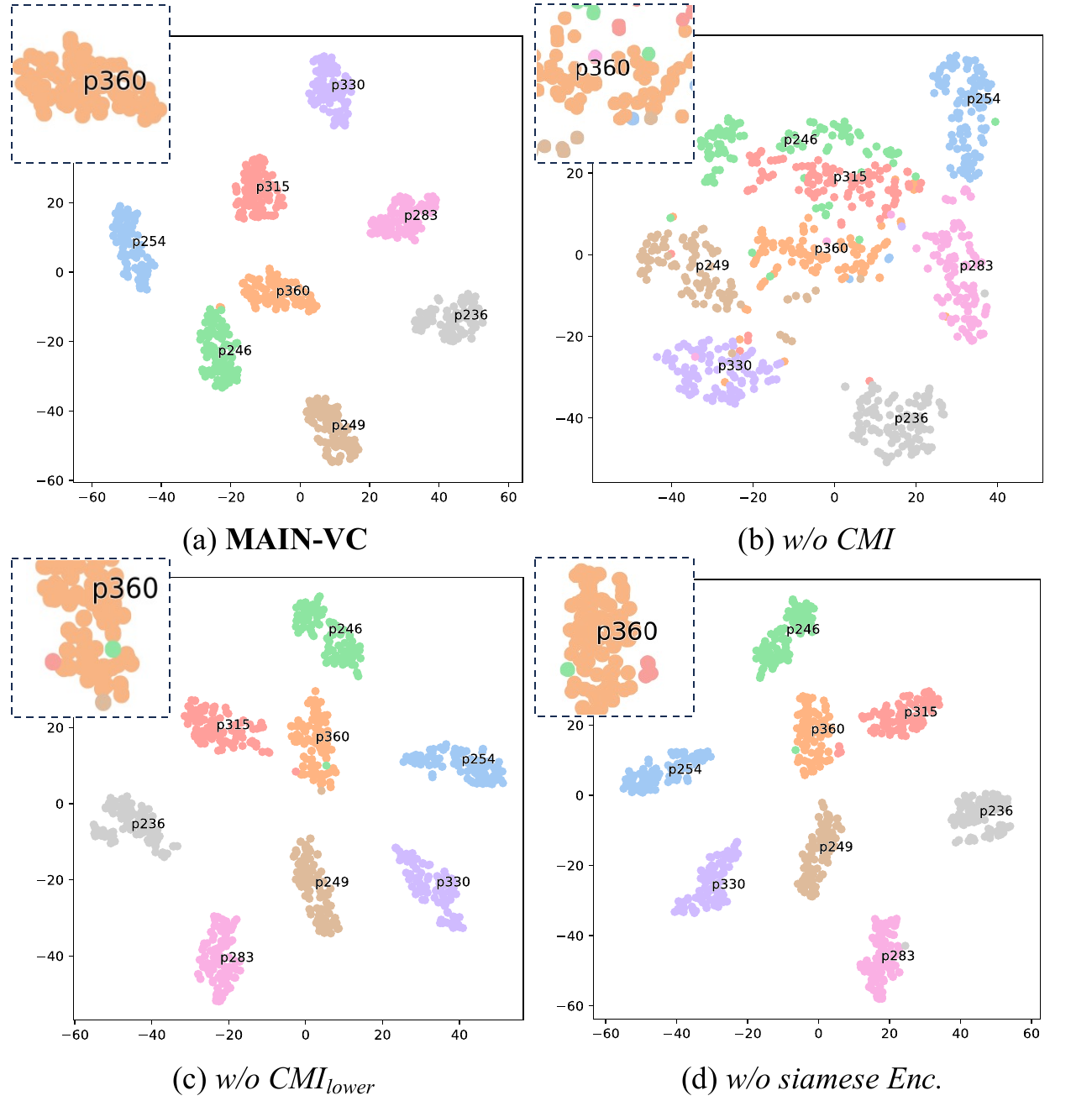}
\setlength{\abovecaptionskip}{0.4cm}
\caption{The visualization of speaker representations extracted
from 8 unseen speakers' utterances (4 males and 4 females, 100 utterances from each speaker).}
\label{fig:tsne}
\end{figure}

To visually assess the disentanglement capability of each model, the scatter diagrams of speaker representations via t-SNE \cite{van2008viual} are shown in Fig. \ref{fig:tsne}. The inclusion of CMI and the Siamese structure in SILM results in more clustered representations of the same speaker, while the removal of the lower bound in CMI leads to looser clustering and inaccurate speaker representation (erroneous clustering in the detail plots), indicating a deterioration in disentangling performance. So SILM and CMI facilitate \textsc{MAIN-VC} to achieve better disentanglement capability and conversion performance. 

\section{Conclusion}
\label{sec:conclusion}

In this paper, we propose a lightweight voice conversion model to address the challenges of VC. The designed speaker information learning module and constrained mutual information estimator effectively enhance the disentanglement capability and improve the conversion quality. From another aspect, with the facilitation of the introduced modules, the disentangling capability of the encoders no longer relies on heavy network structure, so we implement model lightweighting while ensuring the model's performance. The experimental results show that \textsc{MAIN-VC} can complete one-shot voice conversion task and achieves comparable or even better results in various VC scenarios compared to existing methods, while maintaining a lightweight structure.

\section{Acknowledgement}
Supported by the Key Research and Development Program of Guangdong Province (grant No. 2021B0101400003) and the corresponding author is Xulong Zhang (\texttt{zhangxulong@ieee.org}).

\clearpage
\newpage

\bibliographystyle{IEEEtran.bst}
\bibliography{refs.bib}

\begin{thebibliography}{10}
\providecommand{\url}[1]{#1}
\csname url@samestyle\endcsname
\providecommand{\newblock}{\relax}
\providecommand{\bibinfo}[2]{#2}
\providecommand{\BIBentrySTDinterwordspacing}{\spaceskip=0pt\relax}
\providecommand{\BIBentryALTinterwordstretchfactor}{4}
\providecommand{\BIBentryALTinterwordspacing}{\spaceskip=\fontdimen2\font plus
\BIBentryALTinterwordstretchfactor\fontdimen3\font minus \fontdimen4\font\relax}
\providecommand{\BIBforeignlanguage}[2]{{%
\expandafter\ifx\csname l@#1\endcsname\relax
\typeout{** WARNING: IEEEtran.bst: No hyphenation pattern has been}%
\typeout{** loaded for the language `#1'. Using the pattern for}%
\typeout{** the default language instead.}%
\else
\language=\csname l@#1\endcsname
\fi
#2}}
\providecommand{\BIBdecl}{\relax}
\BIBdecl

\bibitem{Yuan/2022/Deid}
R.~Yuan, Y.~Wu, J.~Li, and J.~Kim, ``Deid-vc: Speaker de-identification via zero-shot pseudo voice conversion,'' in \emph{the 23rd Annual Conference of the International Speech Communication Association}, 2022, pp. 2593--2597.

\bibitem{Lal2020evaluate}
B.~M.~L. Srivastava, N.~Vauquier, M.~Sahidullah, A.~Bellet, M.~Tommasi, and E.~Vincent, ``Evaluating voice conversion-based privacy protection against informed attackers,'' in \emph{{IEEE} International Conference on Acoustics, Speech and Signal Processing}, 2020, pp. 2802--2806.

\bibitem{Sisman/2021/An}
B.~Sisman, J.~Yamagishi, S.~King, and H.~Li, ``An overview of voice conversion and its challenges: From statistical modeling to deep learning,'' \emph{{IEEE/ACM} Transactions on Audio, Speech, and Language Processing}, vol.~29, pp. 132--157, 2021.

\bibitem{Kaneko/2018/CycleGAN-VC}
T.~Kaneko and H.~Kameoka, ``Cyclegan-vc: Non-parallel voice conversion using cycle-consistent adversarial networks,'' in \emph{the 26th European Signal Processing Conference}, 2018, pp. 2100--2104.

\bibitem{Kaneko/2021/StarGAN-VC2}
T.~Kaneko, H.~Kameoka, K.~Tanaka, and N.~Hojo, ``Stargan-vc2: Rethinking conditional methods for stargan-based voice conversion,'' in \emph{the 20th Annual Conference of the International Speech Communication Association}, 2019, pp. 679--683.

\bibitem{Li/2021/star}
Y.~A. Li, A.~Zare, and N.~Mesgarani, ``Starganv2-vc: {A} diverse, unsupervised, non-parallel framework for natural-sounding voice conversion,'' in \emph{the 22nd Annual Conference of the International Speech Communication Association}, 2021, pp. 1349--1353.

\bibitem{Chen/2022/effi}
M.~{Chen}, Y.~{Zhou}, H.~{Huang}, and T.~{Hain}, ``{Efficient non-autoregressive gan voice conversion using vqwav2vec features and dynamic convolution},'' \emph{arXiv preprints arXiv:2203.17172}, 2022.

\bibitem{luong2021many}
M.~Luong and V.~Tran, ``Many-to-many voice conversion based feature disentanglement using variational autoencoder,'' in \emph{the 22nd Annual Conference of the International Speech Communication Association}, 2021, pp. 851--855.

\bibitem{xiao2022dgc}
R.~Xiao, H.~Zhang, and Y.~Lin, ``Dgc-vector: {A} new speaker embedding for zero-shot voice conversion,'' in \emph{{IEEE} International Conference on Acoustics, Speech and Signal Processing}, 2022, pp. 6547--6551.

\bibitem{liu2023automatic}
Z.~Liu, S.~Wang, and N.~Chen, ``Automatic speech disentanglement for voice conversion using rank module and speech augmentation,'' in \emph{the 24th Annual Conference of the International Speech Communication Association}, 2023, pp. 2298--2302.

\bibitem{Qian/2019/AutoVC}
K.~Qian, Y.~Zhang, S.~Chang, X.~Yang, and M.~Hasegawa{-}Johnson, ``Autovc: Zero-shot voice style transfer with only autoencoder loss,'' in \emph{the 36th International Conference on Machine Learning}, vol.~97, 2019, pp. 5210--5219.

\bibitem{Chou/2019/One-Shot}
J.~Chou and H.~Lee, ``One-shot voice conversion by separating speaker and content representations with instance normalization,'' in \emph{the 20th Annual Conference of the International Speech Communication Association}, 2019, pp. 664--668.

\bibitem{Wu/2020/One-shot}
D.~Wu and H.~Lee, ``One-shot voice conversion by vector quantization,'' in \emph{{IEEE} International Conference on Acoustics, Speech and Signal Processing}, 2020, pp. 7734--7738.

\bibitem{yuan2021improving}
S.~Yuan, P.~Cheng, R.~Zhang, W.~Hao, Z.~Gan, and L.~Carin, ``Improving zero-shot voice style transfer via disentangled representation learning,'' in \emph{the 9th International Conference on Learning Representations}, 2021.

\bibitem{Qian/2020/Unsupervised}
K.~Qian, Y.~Zhang, S.~Chang, M.~Hasegawa{-}Johnson, and D.~D. Cox, ``Unsupervised speech decomposition via triple information bottleneck,'' in \emph{the 37th International Conference on Machine Learning}, vol. 119, 2020, pp. 7836--7846.

\bibitem{liu2023disentangling}
T.~Liu, K.~A. Lee, Q.~Wang, and H.~Li, ``Disentangling voice and content with self-supervision for speaker recognition,'' in \emph{the 37th Conference on Neural Information Processing Systems}, 2023.

\bibitem{mun2023eend}
S.~H. Mun, M.~H. Han, C.~Moon, and N.~S. Kim, ``Eend-demux: End-to-end neural speaker diarization via demultiplexed speaker embeddings,'' \emph{arXiv preprint arXiv:2312.06065}, 2023.

\bibitem{latif2021survey}
S.~Latif, R.~Rana, S.~Khalifa, R.~Jurdak, J.~Qadir, and B.~Schuller, ``Survey of deep representation learning for speech emotion recognition,'' \emph{IEEE Transactions on Affective Computing}, vol.~14, no.~2, pp. 1634--1654, 2021.

\bibitem{hou2022learning}
N.~Hou, C.~Xu, E.~S. Chng, and H.~Li, ``Learning disentangled feature representations for speech enhancement via adversarial training,'' in \emph{{IEEE} International Conference on Acoustics, Speech and Signal Processing}, 2021, pp. 666--670.

\bibitem{wang2023generalizable}
W.~Wang, Y.~Song, and S.~Jha, ``Generalizable zero-shot speaker adaptive speech synthesis with disentangled representations,'' in \emph{the 24th Annual Conference of the International Speech Communication Association}, 2023.

\bibitem{Chan/2022/SpeechSplit2}
C.~H. Chan, K.~Qian, Y.~Zhang, and M.~Hasegawa{-}Johnson, ``Speechsplit2.0: Unsupervised speech disentanglement for voice conversion without tuning autoencoder bottlenecks,'' in \emph{{IEEE} International Conference on Acoustics, Speech and Signal Processing}, 2022, pp. 6332--6336.

\bibitem{Yen/2021/Again}
Y.~Chen, D.~Wu, T.~Wu, and H.~Lee, ``Again-vc: {A} one-shot voice conversion using activation guidance and adaptive instance normalization,'' in \emph{{IEEE} International Conference on Acoustics, Speech and Signal Processing}, 2021, pp. 5954--5958.

\bibitem{Wu/2020/VQVCPlus}
D.~Wu, Y.~Chen, and H.~Lee, ``Vqvc+: One-shot voice conversion by vector quantization and u-net architecture,'' in \emph{the 21st Annual Conference of the International Speech Communication Association}, 2020, pp. 4691--4695.

\bibitem{Wang/2021/Vqmivc}
D.~Wang, L.~Deng, Y.~T. Yeung, X.~Chen, X.~Liu, and H.~Meng, ``Vqmivc: Vector quantization and mutual information-based unsupervised speech representation disentanglement for one-shot voice conversion,'' in \emph{the 22nd Annual Conference of the International Speech Communication Association}, 2021, pp. 1344--1348.

\bibitem{Yang/2022/Streamable}
H.~Yang, L.~Deng, Y.~T. Yeung, N.~Zheng, and Y.~Xu, ``Streamable speech representation disentanglement and multi-level prosody modeling for live one-shot voice conversion,'' in \emph{the 23rd Annual Conference of the International Speech Communication Association}, 2022, pp. 2578--2582.

\bibitem{Tang/2022/avqvc}
H.~Tang, X.~Zhang, J.~Wang, N.~Cheng, and J.~Xiao, ``Avqvc: One-shot voice conversion by vector quantization with applying contrastive learning,'' in \emph{{IEEE} International Conference on Acoustics, Speech and Signal Processing}, 2022, pp. 4613--4617.

\bibitem{ishmael/2018/mine}
M.~I. Belghazi, A.~Baratin, S.~Rajeshwar, S.~Ozair, Y.~Bengio, A.~Courville, and D.~Hjelm, ``Mutual information neural estimation,'' in \emph{the 35th International Conference on Machine Learning}, ser. Proceedings of Machine Learning Research, vol.~80, 2018, pp. 531--540.

\bibitem{oord2018representation}
A.~v.~d. Oord, Y.~Li, and O.~Vinyals, ``Representation learning with contrastive predictive coding,'' \emph{arXiv preprint arXiv:1807.03748}, 2018.

\bibitem{gutmann2010noise}
M.~Gutmann and A.~Hyv{\"a}rinen, ``Noise-contrastive estimation: A new estimation principle for unnormalized statistical models,'' in \emph{the 13th International Conference on Artificial Intelligence and Statistics}, 2010, pp. 297--304.

\bibitem{ben2019on}
B.~Poole, S.~Ozair, A.~van~den Oord, A.~A. Alemi, and G.~Tucker, ``On variational bounds of mutual information,'' in \emph{the 36th International Conference on Machine Learning}, ser. Proceedings of Machine Learning Research, vol.~97, 2019, pp. 5171--5180.

\bibitem{Cheng/2020/club}
P.~Cheng, W.~Hao, S.~Dai, J.~Liu, Z.~Gan, and L.~Carin, ``Club: {A} contrastive log-ratio upper bound of mutual information,'' in \emph{the 37th International Conference on Machine Learning}, ser. Proceedings of Machine Learning Research, vol. 119, 2020, pp. 1779--1788.

\bibitem{Huang/2017/Arbitrary}
X.~Huang and S.~J. Belongie, ``Arbitrary style transfer in real-time with adaptive instance normalization,'' in \emph{{IEEE} International Conference on Computer Vision}, 2017, pp. 1510--1519.

\bibitem{Liang/2018/Deeplab}
L.~Chen, G.~Papandreou, I.~Kokkinos, K.~Murphy, and A.~L. Yuille, ``Deeplab: Semantic image segmentation with deep convolutional nets, atrous convolution, and fully connected crfs,'' \emph{{IEEE} Transactions on Pattern Analysis and Machine Intelligence}, vol.~40, no.~4, pp. 834--848, 2018.

\bibitem{fisher2017dilated}
F.~Yu, V.~Koltun, and T.~A. Funkhouser, ``Dilated residual networks,'' in \emph{{IEEE} Conference on Computer Vision and Pattern Recognition}, 2017, pp. 636--644.

\bibitem{vctk2016}
J.~Yamagishi, C.~Veaux, and K.~MacDonald, ``Cstr vctk corpus: English multi-speaker corpus for cstr voice cloning toolkit,'' 2016.

\bibitem{kingma2015adam}
D.~P. Kingma and J.~Ba, ``Adam: {A} method for stochastic optimization,'' in \emph{the 3rd International Conference on Learning Representations}, 2015.

\bibitem{nal2018efficient}
N.~Kalchbrenner, E.~Elsen, K.~Simonyan, S.~Noury, N.~Casagrande, E.~Lockhart, F.~Stimberg, A.~van~den Oord, S.~Dieleman, and K.~Kavukcuoglu, ``Efficient neural audio synthesis,'' in \emph{the 35th International Conference on Machine Learning}, ser. Proceedings of Machine Learning Research, vol.~80, 2018, pp. 2415--2424.

\bibitem{Huang/2023/Limivc}
L.~Huang, T.~Yuan, Y.~Liang, Z.~Chen, C.~Wen, Y.~Xie, J.~Zhang, and D.~Ke, ``Limi-vc: A light weight voice conversion model with mutual information disentanglement,'' in \emph{{IEEE} International Conference on Acoustics, Speech and Signal Processing}, 2023, pp. 1--5.

\bibitem{aishell2018}
J.~Du, X.~Na, X.~Liu, and H.~Bu, ``Aishell-2: Transforming mandarin asr research into industrial scale,'' \emph{CoRR}, vol. abs/1808.10583, 2018.

\bibitem{van2008viual}
L.~van~der Maaten and G.~Hinton, ``Viualizing data using t-sne,'' \emph{Journal of Machine Learning Research}, vol.~9, pp. 2579--2605, 11 2008.

\end{thebibliography}

\end{document}